\documentclass{article}

\usepackage{amssymb,amsmath,amsthm}
\usepackage{bbm}
\usepackage{graphicx}
\usepackage{MnSymbol}

\newtheorem{theorem}{Theorem}
\newtheorem{lemma}{Lemma}
\newtheorem{cor}{Corollary}
\theoremstyle{definition}
\newtheorem{defn}{Definition}

\begin{document}
\sloppy

\title{Zeno Squeezing of Cellular Automata}
\author{Martin Schaller}
\author{Karl Svozil}

\author{
Martin Schaller\thanks{
             Algorithmics,
						 Parkring 10, 1010  Vienna, Austria.
             EMail: martin\_schaller@acm.org}
               \, and
Karl Svozil\thanks{
             Institut f\"ur Theoretische Physik, University of Technology Vienna,
						 Wiedner Hauptstra\ss e 8-10/136, A-1040 Vienna, Austria.
             EMail: svozil@tuwien.ac.at}
}

\maketitle

\begin{abstract}
We have recently  introduced the two new computing models of self-similar cellular automata and self-similar Petri nets.
Self-similar automata result from a progressive, infinite tessellation of space and time.
Self-similar Petri nets consist of a potentially infinite sequence of coupled transitions with ever
increasing firing rates.
Both models are capable of hypercomputations and can, for instance, ``solve'' the halting problem for Turing machines.
We survey the main definitions and propositions and add new results regarding the indeterminism of
self-similar cellular automata.
\end{abstract}

\section{Introduction}

Self-similar cellular automata are closely related to cellular automata, a class of dynamical systems
characterized by discreteness in space, time, state values, determinism,
and local interaction (see e.g.,~\cite{gutowitz91}).
A cellular automaton is an infinite lattice of finite automata, each linked with its neighboring automata,
whose underlying space-time structure results from a uniform tessellation of space and time.
In contrast, the underlying space-time structure of a self-similar automaton
is based on a progressive tessellation of space and time, the very same tessellation
that Zeno considered in his paradox of the runner that cannot reach the end of a racecourse
(see e.g.,~\cite{salmon-01}).
Whereas all cells in a one-dimensional cellular automaton are updated synchronously, a cell in
a self-similar cellular automaton is updated twice as often as its left neighbor.
On the one hand, this modification results in completely new capabilities; for instance, there exist
self-similar cellular automata that are capable of hypercomputing.
On the other hand, new paradoxes arise; for instance, the evolution of a self-similar
cellular automaton that involves an infinite number of steps might lead to indeterminism.

The carry-over of the self-similar cellular automaton model to the theory of Petri nets
(see, e.g.,~\cite{Murata89}) yields self-similar Petri nets.
They are equivalent to self-similar cellular automata for a finite number of calculation steps, but differ in the infinite case.
Self-similar Petri nets avoid the indeterminism of self-similar cellular automata by
halting in the infinite case.

There are several aspects that make both
self-similar cellular automata as well as self-similar Petri nets interesting.
Both are extending Zeno's original paradox, leading to a new class of supertasks (see e.g.,~\cite{en96}).

Another notable aspect is the hypercomputing capabilities of both models,
both of them are capable of working as  right-accelerated Turing machines \cite{2008-sica},
which is a subclass of accelerated Turing machines (see e.g.,~\cite{ord-2006}).
Both computing models result from a composition of very basic building blocks --- either finite automata or
Petri net transitions --- and thus demonstrate that it is at least conceptually possible to
build hypercomputers based on those simple building blocks.

Since the two models differ in the infinite limit, the parallelism of both models brings a new twist to Zeno's paradox,
raising new questions about causality and the ontological structure of space and time.

The physical plausibility of accelerated Turing machines, supertasks, and Zeno-like processes,
is discussed elsewhere (see, e.g.,~\cite{sv-aut-rev}).
Originally conceived as a means to demonstrate self-reproduction capabilities in a universal computing
environment by von Neumann~\cite{v-neumann-66},
the idea of perceiving the physical universe as cellular automaton goes back to Zuse \cite{Zuse:67} and was
developed further  by other researchers~\cite{fredkin,toffoli-margolus-90,wolfram-2002}.
Cellular automata based on other tessellations than the uniform grid  were studied in~\cite{margenstern99}.
Hypercomputing is a fast growing field (see, e.g.,~\cite{ord-2006}),
despite criticicism related to the methodology and the classification of
 what should be considered a valid computing process ~\cite{Davis-2004,Davis-2006,potgieter-06}.

In particular, Newtonian mechanics facilitates the construction of hypercomputers.
Davies~\cite{Davies01} describes in some detail how to build an accelerated Turing machine within a Newtonian universe.
For other approaches within Newtonian mechanics, see~\cite{beggs-tucker-06,svozil-2007-cestial}.

As already mentioned, both self-similar cellular automata as well as self-similar Petri nets have been introduced in \cite{2008-sica}.
We survey the main definitions and results and add some further properties of self-similar cellular automata.

The article is organized as follows.
Section \ref{sec:ssca} defines self-similar cellular automata and presents the basic properties.
Section \ref{sec:hypercomputer} is devoted to the construction of a hypercomputer based on
self-similar cellular automata.
Self-similar Petri nets are presented in section \ref{sec:petri}.
This model features a step-to-step equivalence to self-similar cellular automata for finite computations, but halts
in the infinite case.
The final section contains some concluding remarks and gives some directions for future research.

\section{Self-Similar Cellular Automata}
\label{sec:ssca}

\subsection{Basic Definitions}

\begin{figure}
\begin{center}
\scalebox{1.0}{\includegraphics{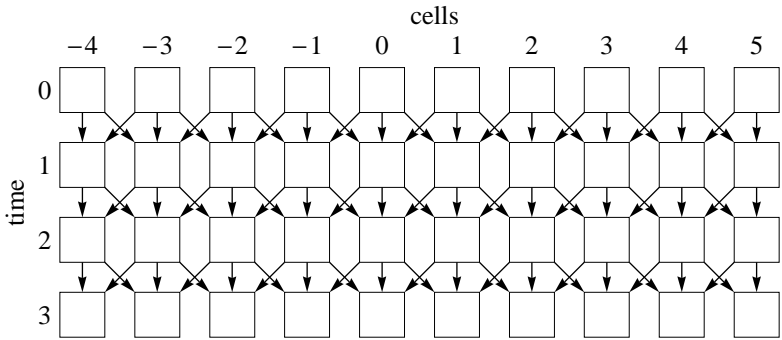}}
\caption{\label{fig:ca-evo} Evolution of a cellular automaton.}
\end{center}
\end{figure}

The underlying structure of a cellular automaton results from a uniform tessellation of space and time.
Fig.~\ref{fig:ca-evo} depicts the evolution of a cellular automaton.
In contrast, self-similar automata result from a progressive tessellation of space and time.
A self-similar cellular automaton operates as a cellular automaton on a one-dimensional lattice containing an infinite number of cells.
Moreover, the cell size and the time  between two updates of the same cell vary depending on the position of the cell in the lattice.
Cell $j$ has size $1/2^j$ and the time between two updates is proportional to the cell size.

One natural way to embed the lattice into $\mathbb{R}$ is the mapping
$j \mapsto 2 - 1/2^{j-1}$ that gives the start point of cell $j$.
Then, the whole lattice maps to  $(-\infty, 2)$, whereby cell $0$ occupies the
unit interval $[0,1)$.

\begin{figure}
\begin{center}
\scalebox{1.0}{\includegraphics{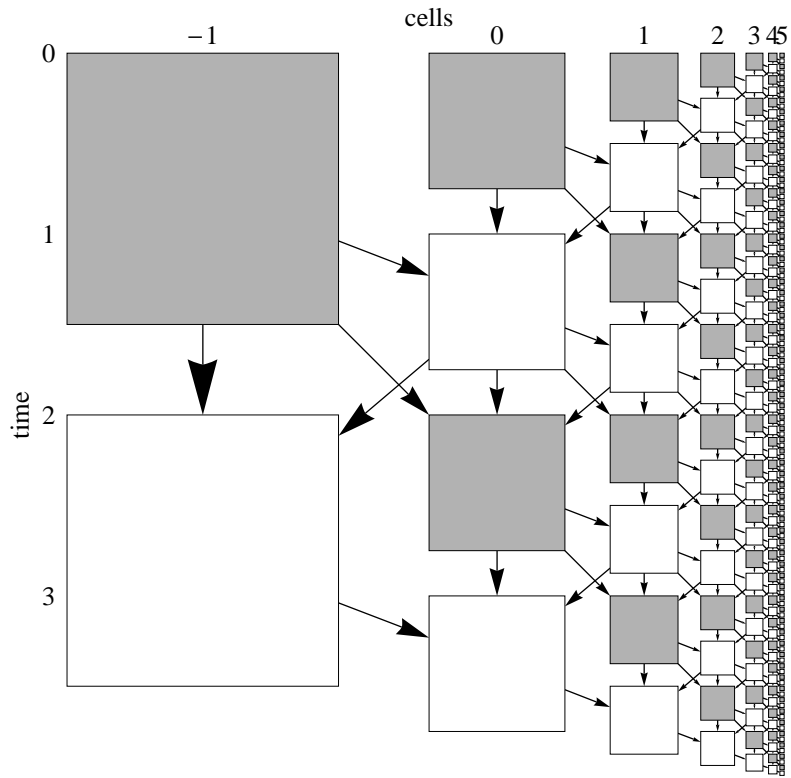}}
\caption{\label{fig:ssca-evo} Evolution of a self-similar cellular automaton.}
\end{center}
\end{figure}

Fig.~\ref{fig:ssca-evo} depicts the evolution of a self-similar cellular automaton in contradistiction to Fig.~\ref{fig:ca-evo}.
Informally speaking, a self-similar cellular automaton features scale-invariance and self-similarity rather
than homogeneity in space and time.

In what follows, we present the formal definition and the description of the update rule.

\begin{defn}
\label{def-ssca}
A self-similar cellular automaton is a tuple $A = (S, f_c, f_d)$, where $S$ is a finite set of states, and
$f_c$ and $f_d$ together represent the local rule, both functions from $S^3$ to $S$.
\end{defn}

Each cell is in a state of the state set $S$.
The state of cell $j$ is updated at times $k / 2^{j}$, where $k$ is an integer.
The cell assumes its new state at time $k / 2^{j}$ and stays in this state until $(k + 1)/2^{j}$, where
the next state change occurs.
The cycle times of a cell are the time intervals from one state transition to the next one, thus,
for cell $j$ these are the half-open intervals $[k/2^j,(k+1)/2^j)$.
This time scheduling implies that the left neighbor cell $j-1$ cycles half as fast, and the right neighbor cell $j+1$ cycles twice as fast as
the cell $j$.
At any given time, the configuration of the automaton is a mapping $c : \mathbb{Z} \rightarrow S$ that specifies the state of all cells.
We denote the state of cell $j$ at time $t$ by $c_j(t)$ and the configuration at $t$ by $c(t)$.

The state of a cell $j$ depends on the last state of the cell itself, and the last states of its left and right neighbor cell.
For notational convenience, we introduce time operators that express the temporal dependencies of a cell.
To this end, we make use of interval arithmetic.
For a scalar $\lambda \in \mathbb{R}$ and a (half-open) interval $[x,y) \subset \mathbb{R}$ set:
$\lambda + [x,y) = [\lambda + x, \lambda + y)$ and $\lambda [x,y) = [\lambda x, \lambda y)$.
We denote the unit interval $[0,1)$ by $\mathbbm{1}$.

If $T=(k + \mathbbm{1})/2^j$ specifies a cycle of cell $j$,
$T_\swarrow= (\lfloor \frac{k-1}{2} \rfloor + \mathbbm{1})/2^{j-1}$ denotes the last cycle of cell $c_{j-1}$,
$T_\downarrow= (k - 1 + \mathbbm{1})/2^j$ the last cycle of cell $c_j$, and
$T_\searrow =  (2k-1 + \mathbbm{1})/2^{j+1}$ the last cycle of cell $c_{j+1}$, respectively, that started before $k/2^j$.
The operator $\downarrow$ is a bijection of the set $\{(k + \mathbbm{1})/2^j| k \in \mathbb{Z} \}$, and we denote by $\uparrow$ its inverse.

The transition of cell $j$ occurs every second time at the times $2k / 2^{j} = k / 2^{j-1}$ synchronously with its left neighbor transition.
A transition of this kind is called $\emph{coupled}$, otherwise it is called $\emph{decoupled}$.
The predicate $\mathit{coupled}((k + \mathbbm{1})/2^j)$ is true if and only if the transition of the $j$-th cell at time $k / 2^{j}$ is coupled, thus,
if and only if $k$ is even.
Cells that have a state resulting from a coupled transition are filled gray in Fig.~\ref{fig:ssca-evo},
the cells that have a state resulting from a decoupled transition are filled white.

The self-similar cellular automaton evolves according to the following update rule.
If $T=(k + \mathbbm{1})/ 2^{j}$ is a cycle of cell $j$, the state  $c_j$ in this interval, formally described by the state function $c_j(T)$, is given by
\begin{equation}
c_j(T) = \left\{
\begin{array}{l}
f_{c}(
	c_{j - 1}(T_\swarrow),
	c_j(T_\downarrow),
	c_{j + 1}(T_\searrow)
)  \mbox{  if $\mathit{coupled}(T)$;} \\
f_{d}(
	c_{j - 1}(T_\swarrow),
	c_j(T_\downarrow),
	c_{j + 1}(T_\searrow)
) \mbox{  if $\neg\mathit{coupled}(T)$.} \\
\end{array}
\right.
\end{equation}
For any time point $t$ and any integer $j$ there exists a unique interval $T=(k + \mathbbm{1})/ 2^{j}$ such that $t \in T$.
This allows us to set $c_j(t) = c_j(T)$.

We remark that only one local rule function is necessary instead of two rule functions $f_c$ and $f_d$,
if an additional flag is added to each state that is toggled for each transition.
For the applications considered later on, the update rule given above is more compact and concise.

\subsection{Indeterminism}

The evolution of a self-similar cellular automaton might become indeterministic.
In what follows we present an example.
Consider the self-similar cellular automaton $A = (\{0,1\}, f_c, f_d)$, where $f_c$ and $f_d$ represent the left shift:
$f_c(?,?,0) = f_d(?,?,0) = 0$ and $f_c(?,?,1) = f_d(?,?,1) = 1$, where the question mark
denotes an arbitrary state.
Suppose $A$ starts at time $0$, and consider the state of cell $0$
at time $1$.
$c_0(1)$ depends on the state $c_1(1/2)$, which itself depends on $c_2(1/4)$, and so on, leading to an infinite regress.
Both possibilities $c_0(1) = c_1(1/2) = c_2(1/4) = \ldots = 0$ and $c_0(1) = c_1(1/2) = c_2(1/4) = \ldots = 1$
are consistent with the local rule and any initial configuration $c(0)$, proving that
the evolution of $A$ is indeterministic and independent of its initial configuration.

Classifying the evolution as indeterministic raises subtle questions that relate to  Thomson's lamp paradox~\cite{thom:54}.
We take the point of view that each cell is at any time in a given state of the state set, even if the initial configuration
and the update rule do not uniquely determine the state.

That the evolution is not necessarily always indeterministic can be seen be the following simple example. 
Assume that the state set $S$ contains a state $q$ satisfying $f_c(?,q,?) = f_d(?,q,?) = q$. 
If cell $j$ is in state $q$, it will for all times stay in this state.
Furthermore, the state of any cell to the left of cell $j$ is deterministic, since the 
causal chain arising in calculating the state of any of these cells stops at cell $j$ and no infinite regress can occur.
For a more subtle example see subsection \ref{sec:block-trans}.

The following lemma reveals limitations of any deterministic evolution.
\begin{lemma}
\label{prop:det}
The state $c_i(t_2)$ of a cell $i$ of a self-similar cellular automata at time $t_2$ 
that was started at $t_1 < t_2$ with configuration $c(t_1)$ is deterministic if and only if
there exists an index $j$ such that $c_i(t_2)$ depends only on states $c_l(t_1)$ with $l < j$.
\end{lemma}
\begin{proof}
We choose $t_1 = 0$, $t_2 = k$, where $k$ is a positive integer, and investigate
whether the state of cell $0$ in the time interval $T=[k,k+1)$ is uniquely determined
by the deterministic states at time 0, that is the configuration $c(0)$.
The general case follows the same proof pattern.
We express a cycle of cell $i$ at time interval $T$ as pair $(i, T)$.
The set of all possible cycles starting not earlier than time $0$ is then
the set $C = \{(i,k + \mathbbm{1})/ 2^{i}| i,k \in \mathbb{Z} \mbox{ and } k \geq 0\}$.

We define a relation $\prec$ on $C$ by setting $(i_1, T_1) \prec (i_2, T_2)$ if and only if
$i_1 = i_2 - 1$ and $T_1 = {T_2}_\swarrow$, or $i_1 = i_2$ and $T_1 = {T_2}_\downarrow$, or $i_1 = i_2 + 1$ and $T_1 = {T_2}_\searrow$.
We denote the transitive closure of $\prec$ by  $\prec^*$.
This relation expresses the possible causal relationship between two transitions.

The set $P = \{(i,T^\prime) \in C| (i,T^\prime) \prec^* (0,T) \}$, the ``past light cone'' of $(0,T)$, contains $(0,T)$ as well as all cycles that
might have an effect on the state of cell $0$ in time interval $T$.
We form increasing subsets of $P$ by setting $P_j = \{(i,T^\prime) \in P| i < j\}$ for $j \geq 0$.

We call a function $s: P_j \rightarrow S$ a realization of $P_j$ if $s$ is consistent with the update rule of the self-similar
cellular automaton and $s$ matches the initial configuration at time $0$.

If we find a $P_j$ such that all realizations of it lead to the same state of cell $0$ at time interval $T$,
we know that the state is deterministic and depends only on cells of the initial configuration with index less than $j$.
Otherwise, if there is no such $P_j$, there are always two realizations $s_1$ and $s_2$ that lead to different states and which can be extended
arbitrarily to the right, resulting in two different evolutions of the self-similar cellular automata and to two different
states of cell $0$ at time interval $T$.
\end{proof}

For the sake of illustration of the implications of this lemma, consider the following example.
Let $C$ be the set of configurations, either of the form $\ldots00100\ldots$, in which exactly one $1$ with a positive index appears, 
or the configuration $0^\infty$ consisting solely of $0$'s.
Assume that a self-similar automaton is started at time $0$ with a configuration $c$ in $C$.
Choose a time $t > 0$ and let $q$ be the state of cell $0$ at time $t$.
Applying the lemma, we see that there exists no local rule such
that $q$ is either $1$ if and only if $c$ is of the form $\ldots00100\ldots$, or $0$ if and only if $c=0^\infty$.
If $q$ is deterministic there is an index $j$ such that 
$q$ depends only on states of cells at time $0$ with index less than $j$. 
This implies that the configurations $0^\infty$ and $\ldots00100\ldots$, where the index of 1 is greater than $j$,
lead to the same state $q$.

\subsection{Self-similar Cellular Automata with Quiescent State}

The indeterminism of self-similar cellular automata can be restricted by considering the following subclass
which adds a quiescent state to the original concept and allows for grids
that contain only a finite number of cells.

\begin{defn}
A self-similiar cellular automaton with quiescent state is a tuple $A = (S, f_c, f_d, q)$, where
$S$, $f_c$, and $f_d$ are defined as in Def.~\ref{def-ssca}, and
$q$ in $S$ is a distinguished state, the quiescent state,
satisfying $f_c(q,q,q) = f_d(q,q,q) = q$.
\end{defn}

If the automaton has a quiescent state, we allow for finite or half-infinite lattices that start with cell $0$.
The update rule of the automaton is adapted to cope with cells that have no left or right neighbor.
Furthermore we allow the lattice to grow to the right.
If either the left or right neighbor is missing, the state of the missing neighbor is assumed to be the quiescent state.
In case of a finite lattice, consisting of $n+1$ cells $0, 1, \ldots, n$, we allow the lattice to grow,
if the state of the $n$-th cell differs from the quiescent state.
If cell $n$ at time $k / 2^n$ changes to a state, different from the quiescent state, a new cell $n+1$ is added to
the lattice.
This new cell $n+1$ is initialised with the quiescent state and attached to cell $n$.
The first update of this new cell occurs at time $(2k+1) / 2^{n+1}$.

A self-cellular automaton with quiescent state is deterministic if the lattice contains only a finite number of cells.

\subsection{Block Transformations}
\label{sec:block-trans}

If the state set becomes larger, the specification of the values for the local rules $f_c$ and $f_d$ for all possible arguments
is rather lengthy.
Some self-similar cellular automata allow an alternative specification.
A coupled transition of two neighbor cells can perform a simultaneous state change of the two cells.
If the state changes of these two neighbor cells are independent of their other neighbors,
we can specify the state changes as a transformation of one state pair into another.
Let $z_1, z_2, z_1^\prime, z_2^\prime$ be elements in $S$.
We call a mapping of the form $z_1 \: z_2 \mapsto z_1^\prime \: z_2^\prime$ a block transformation.
The block transformation $z_1 \: z_2 \mapsto z_1^\prime \: z_2^\prime$ defines
 a function mapping of the form
$
f_c(x, z_1, z_2) = f_d(x, z_1, z_2) = z_1^\prime
$
and
$
f_c(z_1, z_2,y) = z_2^\prime
$
for all $x, y$ in $S$.
Furthermore, we will also allow block transformations that might be ambiguous for certain configurations.
Consider the block transformations
$z_1 \: z_2 \mapsto z_1^\prime \: z_2^\prime$
and
$z_2 \: z_3 \mapsto z_2^{\prime\prime} \: z_3^\prime$
that might lead to an ambiguity for a configuration that contains $z_1z_2z_3$.
Instead of resolving these ambiguities in a formal way, we will restrict our attention to
configurations that are unambiguous.

Consider the self-similar cellular automaton $A = (S, f_c, f_d)$, where $S$ is the set
$ (\{0,1\} \times \{<,>\}) \cup \{\Box\}$.
If $q \in \{0,1\}$, we write $q_<$ for $(q,<)$, and $q_>$ for $(q,>)$, respectively.
We specify  $f_c$ and $f_d$ by the following block transformations
\begin{equation}
0_>\Box \mapsto 0_<\Box, \: 1_>\Box \mapsto 1_<\Box, \: \Box0_< \mapsto \Box0_>, \: \Box1_< \mapsto \Box1_>;
\end{equation}
\begin{equation}
0_> 0_< \mapsto 0_< 0_>, \:  1_> 0_< \mapsto 0_< 1_>, \: 0_> 1_< \mapsto 1_< 0_>, \mbox{ and } 1_> 1_< \mapsto 1_< 1_>;
\end{equation}
together with the convention, that a cell remains in its previous state, if no block transformation is applicable.
Let $A$ be started with a configuration of the form
$\ldots \Box \Box {q_1}_> {q_2}_< {q_3}_> {q_4}_< \ldots {q_{n-1}}_> {q_{n}}_<  \Box \Box \ldots$, where all $q_i$ are in  $\{0,1\}$.
It is easy to see that the evolution of $A$ is deterministic and that
$A$ runs the one-dimensional billard ball model of Margolus \cite{margolus-billard}.
Furthermore, the construction shows that a self-similar cellular automaton can simulate any 3-site
one-dimensional cellular automaton.

\section{Construction of a Hypercomputer}
\label{sec:hypercomputer}

\subsection{Specification}
\label{sec:hc-spec}

In what follows we will construct a hypercomputer based on a self-similar cellular automaton.
This hypercomputer simulates a Turing machine and is capable of performing infinitely many steps
of the Turing machine in finite time.
We assume the following Turing machine model as described in \cite{hopcroft}.

Formally, a {\em Turing machine}  is a tuple
$M = (Q, \Sigma, \Gamma, \delta, q_0, B, F)$,
where $Q$ is the finite set of states, $\Gamma$ is the finite set of tape symbols,
$\Sigma \subset \Gamma$ is the set of input symbols, $q_0 \in Q$ is the start state,
$B \in \Gamma \backslash \Sigma$ is the blank, and $F \subset Q$ is the set of final states.
The next move function or transition function $\delta$ is a mapping from
$Q \times \Gamma$ to $Q \times \Gamma \times \{L, R\}$, which may be undefined for some arguments.

The Turing machine $M$ works on a tape divided into cells that has a leftmost cell but is infinite to the right.
Let $\delta(q, a) = (p, b, D)$.
One step (or move) of $M$ in state $q$ and the head of $M$ positioned over input symbol $a$
consists of the following actions:
scanning input symbol $a$, replacing symbol $a$ by $b$,
entering state $p$ and moving the head one cell either to the left ($D=L$) or to the right ($D=R$).
In the beginning, $M$ starts in state $q_0$ with a tape that is initialized with an input word $w \in \Sigma^*$,
starting at the leftmost cell, all other cells blank,
and the head of $M$ positioned over the leftmost cell.

Given an arbitrary Turing machine $M$ we construct a self-similar cellular automaton with quiescent state $A_M = (Z, f_c, f_d, \Box)$
that simulates $M$.
The state set $Z$ of $A_M$ is given by
\[
Z = \Gamma \cup (\Gamma \times \{\rightarrow\}) \cup (Q \times \Gamma)
\cup (Q \times \Gamma \times \{\rightarrow\}) \cup
\{\Box, \blacktriangleleft, \lhd, \overrightarrow{\lhd}, \rhd, \rhd_B, \rhd_\blacktriangleleft\}.
\]
We write $\overrightarrow{a}$ for an element $(a, \rightarrow)$ in $\Gamma \times \{\rightarrow\}$,
$\langle q,a \rangle$ for an element
$(q, a)$ in $Q \times \Gamma$, and
$\overrightarrow{\langle q,a \rangle}$ for an element
$(q, a, \rightarrow)$ in $Q \times \Gamma \times \{\rightarrow\}$.
To simulate $M$ on input $w=a_1 \ldots a_n$ in $\Sigma^*$, $n > 1$,
$A_M$ is initialized with the sequence
$\overrightarrow{\lhd} \langle q_0,a_1 \rangle a_2 a_3\ldots a_n\rhd $
starting at cell 0.
If $w=a_1$, $A_M$ is initialized with the sequence
$\overrightarrow{\lhd} \langle q_0,a_1 \rangle B\rhd $, and
if $w=\epsilon$, the empty word, $A_M$ is initialized with the sequence
$\overrightarrow{\lhd} \langle q_0,B \rangle B\rhd $.
We denote the initial configuration by $c(w)$.
The computation is started at time 0, i.e., the first state change of cell $k$ occurs at time $1/2^k$.

We specify the local rule, represented by $f_c$ and $f_d$, by the following block transformations, together with the convention
that a cell state remains unchanged, if no block transformation is applicable.
\begin{enumerate}

\item
\emph{Pulse moves to the right.}
Set
\begin{equation}
\overrightarrow{\lhd} \: \langle q, a \rangle \mapsto \lhd \: \overrightarrow{\langle q, a \rangle};
\label{tr:start-state}
\end{equation}
\begin{equation}
\overrightarrow{a} \: b \mapsto a \: \overrightarrow{b};
\label{tr:down}
\end{equation}
\begin{equation}
\overrightarrow{\lhd} \:a \mapsto \lhd \: \overrightarrow{a}.
\label{tr:start}
\end{equation}
If $\delta(q,a) = (p,c,R)$ set
\begin{equation}
\overrightarrow{b} \: \langle q, a \rangle \mapsto b \:
\overrightarrow{\langle q, a \rangle};
\label{tr:down-to-head}
\end{equation}
\begin{equation}
\overrightarrow{\langle q,a \rangle} \: b \mapsto c \:
\overrightarrow{\langle p, b \rangle};
\label{tr:right-2}
\end{equation}
\begin{equation}
\overrightarrow{\langle q,a \rangle} \: \rhd \mapsto \langle q,a \rangle \:
\rhd_B.
\label{tr:down-state-right-delimiter-blank}
\end{equation}
If $\delta(q,a) = (p,c,L)$ set
\begin{equation}
\overrightarrow{b} \: \langle q, a \rangle \mapsto \langle p, b \rangle \:
\overrightarrow{c};
\label{tr:left-1}
\end{equation}
\begin{equation}
\overrightarrow{\langle q,a \rangle} \: b \mapsto \langle q,a \rangle \:
\overrightarrow{b};
\label{tr:left-no-move}
\end{equation}
\begin{equation}
\overrightarrow{\langle q,a \rangle} \: \rhd \mapsto \langle q,a \rangle \:
\rhd_\blacktriangleleft.
\label{tr:down-state-right-delimiter}
\end{equation}
Set
\begin{equation}
\overrightarrow{a} \: \rhd \mapsto a \: \rhd_\blacktriangleleft;
\label{tr:down-a-rhd}
\end{equation}
\begin{equation}
\rhd_B \: \Box \mapsto B \: \rhd_\blacktriangleleft;
\label{tr:new-blank}
\end{equation}
\begin{equation}
\rhd_\blacktriangleleft \: \Box \mapsto \blacktriangleleft \: \rhd.
\label{tr:reflection-right}
\end{equation}

\item
\emph{Pulse moves to the left}.
Set
\begin{equation}
a \: \blacktriangleleft \mapsto \blacktriangleleft \: a;
\label{tr:up}
\end{equation}
\begin{equation}
\langle q,a \rangle \: \blacktriangleleft \mapsto  \blacktriangleleft \:
\langle q,a \rangle;
\label{tr:up-state}
\end{equation}
\begin{equation}
\lhd \: \blacktriangleleft \mapsto \Box \: \overrightarrow{\lhd}.
\label{tr:up-lhd}
\end{equation}
\end{enumerate}

The states $\langle q, a \rangle$ and $\overrightarrow{\langle q, a \rangle}$ act as the head of the Turing machine.
To accelerate the calculation we have to shift the whole tape content of the Turing machine to the right, to
the faster cycling cells, thereby avoiding that the content is spread, i.e., that right tape states move faster than
left tape states.

This can be achieved by synchronizing all state transitions by a pulse ($\overrightarrow{\lhd}, \overrightarrow{a},
\overrightarrow{\langle q,a \rangle}, \rhd_B$, $\rhd_\blacktriangleleft$,
or $\blacktriangleleft$) that zigzags between the two delimiters $\lhd$ and $\rhd$.
Additionally, new blanks are inserted  to the left of the right delimiter whenever the simulated
head of the Turing machine hits the right delimiter and attempts to move to the right.

\subsection{Example}

\begin{figure}
\begin{center}
\renewcommand{\arraystretch}{1}
\begin{tabular}{c|ccccc}
& \multicolumn{5}{c}{ Symbol} \\
State & 0 & 1 & $X$ & $Y$ & $B$ \\ \hline
$q_0$ & $(q_1,X,R)$ & ---         & ---         & $(q_3,Y,R)$ & ---         \\
$q_1$ & $(q_1,0,R)$ & $(q_2,Y,L)$ & ---         & $(q_1,Y,R)$ & ---         \\
$q_2$ & $(q_2,0,L)$ & ---         & $(q_0,X,R)$ & $(q_2,Y,L)$ & ---         \\
$q_3$ & ---         & ---         & ---         & $(q_3,Y,R)$ & $(q_4,B,R)$ \\
$q_4$ & ---         & ---         & ---         & ---         & ---         \\
\end{tabular}
\end{center}
\caption{\label{fig:example-delta}The function $\delta$.}
\end{figure}

We illustrate the working of $A_M$ by a simple example \cite{2008-sica}.
Let $L$ be the formal language consisting of strings with $n$ 0's, followed by $n$ 1's:
$L = \{0^n1^n | n \geq 1\}$.
A Turing machine that accepts this language is given by
$M = (\{q_0, q_1, q_2, q_3, q_4\}, \{0,1\}, \{0,1,X,Y,B\}, \delta, q_0, B, \{q_4\})$ \cite{hopcroft}
with the transition function depicted in Fig.~\ref{fig:example-delta}.
Note that $L$ is a context-free language, but $M$ will serve for demonstration purposes.
The computation of $M$ on input $01$ is given below:
\[
q_001 \vdash Xq_11  \vdash  q_2XY  \vdash  Xq_0Y  \vdash  XYq_3  \vdash XYBq_4.
\]
Fig.~\ref{fig:example-hyper-sca-2} depicts the computation of $A_M$ on the Turing machine input 01, 
showing only the configurations where a state change occurred.
The first column of the table specifies the time in binary base.
$A_M$ performs 4 complete pulse zigzags and enters a final configuration in the fifth one after the Turing machine simulation has reached
the final state $q_4$.

\begin{figure}
\begin{center}
\footnotesize     {
\renewcommand{\arraystretch}{1.0}
\begin{tabular}{r|c@{\hspace{2mm}}c@{}c@{}c@{}c@{}c@{}c@{}c@{}c@{\hspace{3mm}}c@{}c@{}c@{}c@{}c}
   &0 &1 & 2 & 3 & 4 & 5 & 6 & 7 & 8 & 9 \\ \hline
$0.00000000_2$ & $\overrightarrow{\lhd}$ & $\langle q_0,0 \rangle$ & $1$ & $\rhd$ \\
$1.00000000_2$ & $\lhd$ & $\overrightarrow{\langle q_0,0 \rangle}$ & $1$ & $\rhd$  \\
$1.10000000_2$ & $\lhd$ & $X$ & $\overrightarrow{\langle q_1,1 \rangle}$ & $\rhd$  \\
$1.11000000_2$ & $\lhd$ & $X$ & $\langle q_1,1 \rangle$ & $\rhd_\blacktriangleleft$ & \\
$1.11100000_2$ & $\lhd$ & $X$ & $\langle q_1,1 \rangle$ & $\blacktriangleleft$ & $\rhd$ \\
$10.00000000_2$ & $\lhd$ & $X$ & $\blacktriangleleft$ & $\langle q_1,1 \rangle$ & $\rhd$ \\
$10.10000000_2$ & $\lhd$ & $\blacktriangleleft$ & $X$ & $\langle q_1,1 \rangle$ & $\rhd$ \\
$11.00000000_2$ & $\Box$ & $\overrightarrow{\lhd}$ & $X$ & $\langle q_1,1 \rangle$ & $\rhd$ \\
$11.10000000_2$ & $\Box$ & $\lhd$ & $\overrightarrow{X}$ & $\langle q_1,1 \rangle$ & $\rhd$ \\
$11.11000000_2$ & $\Box$ & $\lhd$ & $\langle q_2,X \rangle$ & $\overrightarrow{Y}$ & $\rhd$ \\
$11.11100000_2$ & $\Box$ & $\lhd$ & $\langle q_2,X \rangle$ & $Y$ & $\rhd_\blacktriangleleft$ \\
$11.11110000_2$ & $\Box$ & $\lhd$ & $\langle q_2,X \rangle$ & $Y$ & $\blacktriangleleft$ & $\rhd$ \\
$100.00000000_2$ & $\Box$ & $\lhd$ & $\langle q_2,X \rangle$ & $\blacktriangleleft$ & $Y$ & $\rhd$ \\
$100.01000000_2$ & $\Box$ & $\lhd$ & $\blacktriangleleft$ & $\langle q_2,X \rangle$ & $Y$ & $\rhd$ \\
$100.10000000_2$ & $\Box$ & $\Box$ & $\overrightarrow{\lhd}$ & $\langle q_2,X \rangle$ & $Y$ & $\rhd$ \\
$100.11000000_2$ & $\Box$ & $\Box$ & $\lhd$ & $\overrightarrow{\langle q_2,X \rangle}$ & $Y$ & $\rhd$ \\
$100.11100000_2$ & $\Box$ & $\Box$ & $\lhd$ & $X$ & $\overrightarrow{\langle q_0,Y \rangle}$ & $\rhd$ \\
$100.11110000_2$ & $\Box$ & $\Box$ & $\lhd$ & $X$ & $\langle q_0,Y \rangle$ & $\rhd_B$ \\
$100.11111000_2$ & $\Box$ & $\Box$ & $\lhd$ & $X$ & $\langle q_0,Y \rangle$ & $B$ & $\rhd_\blacktriangleleft$ \\
$100.11111100_2$ & $\Box$ & $\Box$ & $\lhd$ & $X$ & $\langle q_0,Y \rangle$ & $B$ & $\blacktriangleleft$ & $\rhd$ & \\
$101.00000000_2$ & $\Box$ & $\Box$ & $\lhd$ & $X$ & $\langle q_0,Y \rangle$ & $\blacktriangleleft$ & $B$ & $\rhd$ & \\
$101.00010000_2$ & $\Box$ & $\Box$ & $\lhd$ & $X$ & $\blacktriangleleft$ & $\langle q_0,Y \rangle$ & $B$ & $\rhd$ & \\
$101.00100000_2$ & $\Box$ & $\Box$ & $\lhd$ & $\blacktriangleleft$ & $X$ & $\langle q_0,Y \rangle$ & $B$ & $\rhd$ & \\
$101.01000000_2$ & $\Box$ & $\Box$ & $\Box$ & $\overrightarrow{\lhd}$ & $X$ & $\langle q_0,Y \rangle$ & $B$ & $\rhd$ & \\
$101.01100000_2$ & $\Box$ & $\Box$ & $\Box$ & $\lhd$ & $\overrightarrow{X}$ & $\langle q_0,Y \rangle$ & $B$ & $\rhd$ & \\
$101.01110000_2$ & $\Box$ & $\Box$ & $\Box$ & $\lhd$ & $X$ & $\overrightarrow{\langle q_0,Y \rangle}$ & $B$ & $\rhd$ & \\
$101.01111000_2$ & $\Box$ & $\Box$ & $\Box$ & $\lhd$ & $X$ & $Y$ & $\overrightarrow{\langle q_3,B \rangle}$ & $\rhd$ & \\
$101.01111100_2$ & $\Box$ & $\Box$ & $\Box$ & $\lhd$ & $X$ & $Y$ & $\langle q_3,B \rangle$ & $\rhd_B$ & \\
$101.01111110_2$ & $\Box$ & $\Box$ & $\Box$ & $\lhd$ & $X$ & $Y$ & $\langle q_3,B \rangle$ & $B$ & $\rhd_\blacktriangleleft$ \\
$101.01111111_2$ & $\Box$ & $\Box$ & $\Box$ & $\lhd$ & $X$ & $Y$ & $\langle q_3,B \rangle$ & $B$ & $\blacktriangleleft$ & $\rhd$ \\
$101.10000000_2$ & $\Box$ & $\Box$ & $\Box$ & $\lhd$ & $X$ & $Y$ & $\langle q_3,B \rangle$ & $\blacktriangleleft$ & $B$ & $\rhd$ \\
$101.10000100_2$ & $\Box$ & $\Box$ & $\Box$ & $\lhd$ & $X$ & $Y$ & $\blacktriangleleft$ & $\langle q_3,B \rangle$ & $B$ & $\rhd$ \\
$101.10001000_2$ & $\Box$ & $\Box$ & $\Box$ & $\lhd$ & $X$ & $\blacktriangleleft$ & $Y$ & $\langle q_3,B \rangle$ & $B$ & $\rhd$ \\
$101.10010000_2$ & $\Box$ & $\Box$ & $\Box$ & $\lhd$ & $\blacktriangleleft$ & $X$ & $Y$ & $\langle q_3,B \rangle$ & $B$ & $\rhd$ \\
$101.10100000_2$ & $\Box$ & $\Box$ & $\Box$ & $\Box$ & $\overrightarrow{\lhd}$ & $X$ & $Y$ & $\langle q_3,B \rangle$ & $B$ & $\rhd$ \\
$101.10110000_2$ & $\Box$ & $\Box$ & $\Box$ & $\Box$ & $\lhd$ & $\overrightarrow{X}$ & $Y$ & $\langle q_3,B \rangle$ & $B$ & $\rhd$ \\
$101.10111000_2$ & $\Box$ & $\Box$ & $\Box$ & $\Box$ & $\lhd$ & $X$ & $\overrightarrow{Y}$ & $\langle q_3,B \rangle$ & $B$ & $\rhd$ \\
$101.10111100_2$ & $\Box$ & $\Box$ & $\Box$ & $\Box$ & $\lhd$ & $X$ & $Y$ & $\overrightarrow{\langle q_3,B \rangle}$ & $B$ & $\rhd$ \\
$101.10111110_2$ & $\Box$ & $\Box$ & $\Box$ & $\Box$ & $\lhd$ & $X$ & $Y$ & $B$ & $\overrightarrow{\langle q_4,B \rangle}$ & $\rhd$ \\
\end{tabular}
}
\end{center}
\caption{\label{fig:example-hyper-sca-2}A computation of $A_M$ on input $01$ \cite{2008-sica}.}
\end{figure}

\subsection{Results}

As one can see in Fig.~\ref{fig:example-hyper-sca-2}, a ``zigzag'' pulse that goes from a configuration
containing state $\overrightarrow{\lhd}$ to the next one that contains the same state takes three
cycles of the cell where the pulse has started.
Afterwards, the whole tape content is shifted one cell to the right.
Thus, the whole simulation takes no longer than $3 + 3/2 + 3/4 + \ldots = 6$ time units.
Furthermore, each pulse zigzag performs at least one step of the simulated Turing machine.
Taking these facts together, the following theorem is motivated that was proved in \cite{2008-sica}.

\begin{theorem}
Let $M$ be a Turing machine, $w$ an input word of $M$, and $A_M$ the self-similar automaton given above that simulates $M$,
initialised with $c(w)$.
If $M$ halts on $w$, then $A_M$ enters a final configuration in a time less than 6 cycles of cell $0$.
If $M$ does not halt, $A_M$ enters after 6 cycles of cell $0$ the quiescent configuration $\Box^\infty$.
\end{theorem}

If we choose for $M$ a universal Turing machine, we obtain the following result, which proves that $A_M$ is a hypercomputer
for certain Turing machines $M$.
\begin{cor}
Let $M_U$ be a universal Turing machine. Then $A_{M_U}$ solves the halting problem for Turing machines.
\end{cor}

We imagine that an operator initialises the first cells of the self-similar automaton with the input of the calculation.
Ideally, in case that the simulated Turing machine has halted, the self-similar automaton should propagate this fact
back to the left cells.
But by lemma \ref{prop:det} we know that there is no deterministic way to do this.
Therefore the operator would have to scan a possible infinite numbers of cells to decide whether the Turing machine has
halted or not.

\section{Self-similar Petri Nets}
\label{sec:petri}

\begin{figure}
\begin{center}
\includegraphics[scale=0.45]{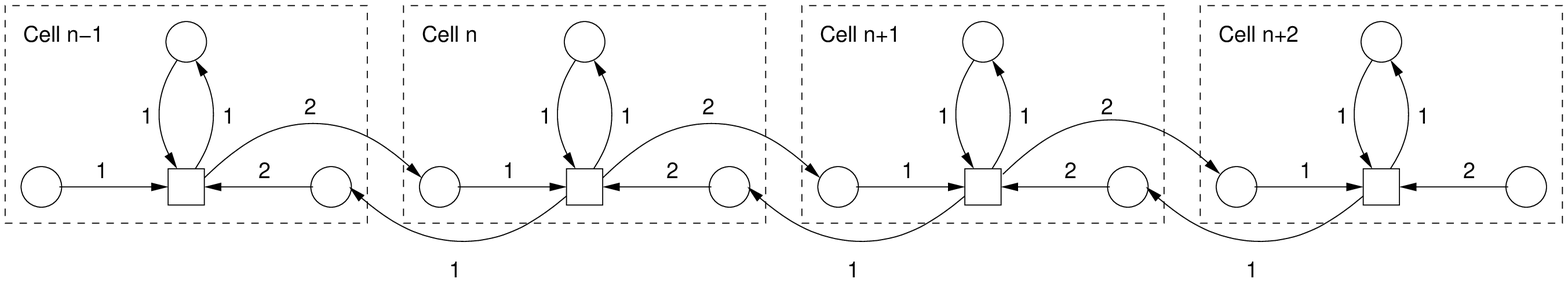}
\caption{\label{fig:petri} Underlying graph of a self-similar Petri net \cite{2008-sica}.}
\end{center}
\end{figure}

Self-similar Petri nets result from carrying over the self-similar cellular automaton model to the theory of Petri nets.
We refer to \cite{Murata89} for a concise introduction to Petri net theory, here we give only a very short
summary to settle the terminology.

The underlying graph of a Petri net is a directed, weighted, bipartite graph consisting of two kind of nodes,
called transitions and places.
Fig.~\ref{fig:petri} depicts the underlying graph of a self-similar Petri net, drawing
transitions as boxes and places as circles.
A place that has an arc to a transition is an input place of this transition,
if the arc is from the transition to the place, the place is an output place.
Arcs are labeled with their weights.

Places hold so-called tokens.
A marking assigns to each place a number, the number of tokens in this place.
The marking in a Petri net is changed according to the following transition (firing) rule:
\begin{enumerate}
\item A transition $t$ is enabled if each input place $p$ of $t$ is marked with at
least $w(p,t)$ tokens, where $w(p,t)$ is the weight of the arc from $p$ to $t$.
\item An enabled transition $t$ may fire. A firing removes $w(p,t)$ tokens from each input place $p$,
and adds $w(t,p)$ tokens to each output place $p$ of $t$, where $w(t,p)$ is the weight
of the arc from $t$ to $p$.
\end{enumerate}

Self-similar Petri nets are both colored Petri nets and marked graphs.
The first says that the tokens of the Petri net carry values and that
the firing rule is adapted such that the value of an output token is determined by the values of the input tokens.
The latter says that each place is the input place and the output place of at most one transition, which
makes the Petri net deterministic.

We will informally describe how the concepts of self-similar cellular automata are mapped to self-similar Petri nets,
for a formal treatment we refer to \cite{2008-sica}.
The states of a self-similar cellular automaton are mapped to the values of the tokens.
The transition of the self-similar Petri net uses the values of the input tokens to calculate
the value of the output tokens according to the local rules $f_c$ and $f_d$ that are carried over from self-similar cellular automata.

A firing of cell $n$ consumes two tokens of cell $n+1$ and puts two new tokens in the input place of cell $n+1$.
Since cell $n+1$ consumes per firing only one token from cell $n$, and puts only one token in the input place of cell $n$,
cell $n+1$ must fire twice before cell $n$ can fire again.
As we can see, the doubling of cycles from one cell to its right neighbor works now by a synchronisation mechanism without reference to
an external clock.

In analogy to self-similar cellular automata with quiescent state, a self-similar Petri net
is started with a finite number of cells and is allowed to grow to the right, whenever the rightmost cell
calculates a token value different from the quiescent state.

To ensure the liveness of the self-similar Petri net the left- and rightmost cells 
obey the the following boundary conditions.
Each firing of the leftmost cell puts one token in its left input place, each firing of the rightmost cell 
puts two tokens in its right input place.

If the self-similar Petri net is started with a certain marking and proper token values it can be shown
that self-similar Petri nets and self-similar cellular automata feature a step-by-step equivalence for calculations that
involve only a finite number of steps.

Self-similar Petri nets work without any reference to an external clock, but it is possible to impose a time scheduling
leading to timed self-similar Petri nets.
If we require that transition $n$ always fires when it is enabled and that the firing process, which includes
the consumption and production of tokens, takes no longer than $1/2^n$ time units, we obtain
the same time model as for self-similar cellular automata.

The construction of subsection \ref{sec:hc-spec} can also be applied to timed self-similar Petri nets,
leading to one that simulates a given Turing machine with a given input.

In contrast to self-similar cellular automata, the evolution of a Petri net can stop.
This happens when no transition is enabled. The main result concerning timed self-similar Petri nets is expressed by
the following theorem.
\begin{theorem}
Let $M$ be a Turing machine, $w$ an input word of $M$, and $N_M$ a timed self-similar Petri net that simulates $M$,
initialised with $c(w)$.
If $M$ halts on $w$, then $N_M$ enters a final configuration in a time less than 6 cycles of cell $0$.
If $M$ does not halt, $N_M$ halts after 6 cycles of cell $0$.
\end{theorem}
Again, we refer the reader for details and the proof to \cite{2008-sica}.

\section{Summary}

We have reviewed two recently introduced computing computing models, both based on an infinite,
progressive tessellation of space and time, thereby proving a proposition about the indeterminism of self-similar cellular automata.
Space and time tessellations are the same as imagined in Zeno's paradox of the runner that cannot reach
the end of a racecourse more than 2500 years ago.

Both computing models are capable of hypercomputing, even if they differ in the limit of non-halting Turing machine simulations.
If properly programmed, self-similar cellular automata enter a final quiescent configuration and loop  forever there; if not,
they end up in indeterminism.

The underlying graph of a self-similar Petri net grows to 
infinity, if the simulated Turing machine does not halt.
Since there is no longer a rightmost cell that obeys the boundary condition that guaranteed the liveness 
of the system for the finite case, the self-similar Petri net stops.
Thus, self-similar Petri nets halt if and only if the simulated Turing machine does not halt.

Both models suffer from what we call the response problem.
We imagine an operator that initialises the very first cells of either of the two machines with the input of the
calculation and then starts the machine.
Ideally, after some finite amount of time the operator would obtain an answer that is again written
to the first cells of the machine.
Thus, the response problem is the problem of propagating the final status of the simulated Turing machine,
 which is either ``halt" or ``non-halt," back to the cells with lower index,
say cell 0.

Both models fail to solve the response problem; yet due to different reasons.
If we extend the rules of the self-similar cellular automaton to propagate a response back to the left cells,
the automaton becomes indeterministic.
In contrast, self-similar Petri nets freeze if they run into infinity, thereby eliminating any possibility to
propagate information backward.
The possibility of the (non)existence of other elementary computational models whose Zeno squeezed versions  on the one hand are
capable of hypercomputing, yet on the other hand do not suffer from the response problem, remains an open question.


\end{document}